\documentclass[letter,traditabstract]{aa} % for a referee version
\usepackage{graphicx}
\usepackage{natbib}
\usepackage{txfonts}
\def\igr{IGR\,J17361-4441}
\def\inte{{\em INTEGRAL}}
\def\xmm{{\em XMM-Newton}}

\def\chan{{\em Chandra}}

\def\rxte{{\em RXTE}}
\def\swift{{\em Swift}}

\def \inte {{\em INTEGRAL}}
\def \xmm {{\em XMM--Newton}}
\def \chandra {{$Chandra$}}

\def \hcm {\hbox {\ifmmode $ atom cm$^{-2}\else atom cm$^{-2}$\fi}}

\def \apjl {ApJL}
\def \apjs {ApJS}\def \aap {A\&A}

\begin{document}
   \title{IGRJ17361-4441: a possible new accreting X-ray binary in NGC6388}

   \author{E. Bozzo
          \inst{1}
      \and C. Ferrigno
           \inst{1}
      \and J. Stevens
            \inst{2}
      \and T. M. Belloni
            \inst{3}     
      \and J. Rodriguez
            \inst{4}
      \and P. R. den Hartog
            \inst{5}
      \and A. Papitto
            \inst{6}
      \and I. Kreykenbohm
            \inst{7}
      \and F. Fontani 
            \inst{8}
      \and L. Gibaud 
            \inst{1}
          }

   \institute{ISDC Data Centre for Astrophysics, Chemin d’Ecogia 16,
             CH-1290 Versoix, Switzerland; \email{enrico.bozzo@unige.ch}
              \and 
              CSIRO Astronomy \& Space Science, Australia Telescope National Facility, P.O. Box 76, Epping, NSW 1710, Australia 
              \and  
              INAF, Osservatorio Astronomico di Brera, via E. Bianchi 46, I-23807 Merate (LC), Italy 
              \and
              Laboratoire AIM, CEA/IRFU - CNRS/INSU - Universit\'e Paris Diderot, CEA DSM/IRFU/SAp, F-91191 Gif-sur-Yvette, France
              \and 
              Stanford University HEPL/KIPAC Physics, 382 Via Pueblo Mall Stanford, 94305, USA
              \and
              Dipartimento di Fisica, Universit\'a degli Studi di Cagliari, SP Monserrato-Sestu, KM 0.7, 09042 Monserrato, Italy
              \and
              Dr. Karl Remeis-Sternwarte \& Erlangen Centre for Astroparticel Physics, Sternwartstr. 7, 96049 Bamberg 
              \and
              INAF - Osservatorio Astrofisico di Arcetri, L.go E. Fermi 5, I-50125, Firenze, Italy
              }

   \date{Submitted: 2011 September 5 - Accepted: 2011 September 22}

  \abstract{\igr\ is a newly discovered \inte\ hard X-ray transient, located in the globular cluster NGC6388.
  We report here the results of the X-ray and radio observations performed with \swift,\ \inte,\ \rxte,\ and the 
  Australia Telescope Compact Array (ATCA) after the discovery of the source on 2011 August 11. 
  In the X-ray domain, \igr\ showed virtually constant flux and spectral parameters up to 18~days 
  from the onset of the outburst. The broad-band (0.5-100~keV) spectrum of the source could be reasonably 
  well described by using an absorbed power-law component with a high energy cut-off 
  ($N_{\rm H}$$\simeq$0.8$\times$10$^{22}$~cm$^{-2}$, $\Gamma$$\simeq$0.7-1.0, and $E_{\rm cut}$$\simeq$25~keV) and 
  displayed some evidence of a soft component below $\sim$2~keV.  
  No coherent timing features were found in the \rxte\ data. 
  The ATCA observation did not detect significant radio emission from \igr,\ and provided the most 
  stringent upper limit (rms 14.1~$\mu$Jy at 5.5~GHz) to date on the presence of any radio source close to the NGC6388 
  center of gravity. The improved position of \igr\ in outburst determined from a recent target of opportunity observation with \chan,\ together with the 
  X-ray flux and radio upper limits measured in the direction of the source, argue against its association with the putative intermediate-mass black 
  hole residing in the globular cluster and with the general hypothesis that the \inte\ source is a black hole candidate. 
  \igr\ might be more likely a new X-ray binary hosting an accreting neutron star.
  The ATCA radio non-detection also permits us to derive an upper limit to the mass of the suspected 
  intermediate massive black hole in NGC6388 of $\lesssim$600~$M_{\sun}$. This is a factor of 2.5 lower than the limit 
    reported previously.} 

  \keywords{gamma rays: observations -- X-rays: individuals: IGR J17361-4441}

   \maketitle

\section{Introduction}
\label{sec:intro}

\igr\ was discovered by the IBIS/ISGRI telescope \citep{ubertini03} 
onboard \inte\ \citep{winkler03} on 2011 August 11 \citep{gibaud11}. 
The source was located in the direction of the globular cluster (GC) 
NGC6388 and detected at an X-ray flux of 9.7$\times$10$^{-11}$~erg~cm$^{-2}$~s$^{-1}$ 
(20-100~keV).  
Follow-up observations carried out with \swift\ and \chan\ \citep{ferrigno11b,pooley11} provided  
a first description of the source X-ray emission in the soft energy band (0.3-10~keV),  
a refined source position at RA = 17:36:17.418, Dec = -44:44:05.98 (J2000; the nominal 
\chan\ positional accuracy is 0.6~arcsec, 90\% c.l.) and confirmed its localization in NGC6388. 
In the past, several X-ray observational campaigns have been carried out in the direction of this GC, 
as it is believed to host an intermediate-mass black hole (IMBH) at its center of gravity  
\citep[hereafter COG;][]{lanzoni07}. Here we report on the analysis of all the available 
X-ray data collected during the first 18~days of the outburst observed from \igr,\ and use a  
simultaneous observation in the radio domain with the ATCA telescope to investigate 
the nature of the source.

\section{Data analysis}

\subsection{ \inte\ }
\label{sec:inte}
 
\begin{figure}
  \includegraphics[width=8.4cm]{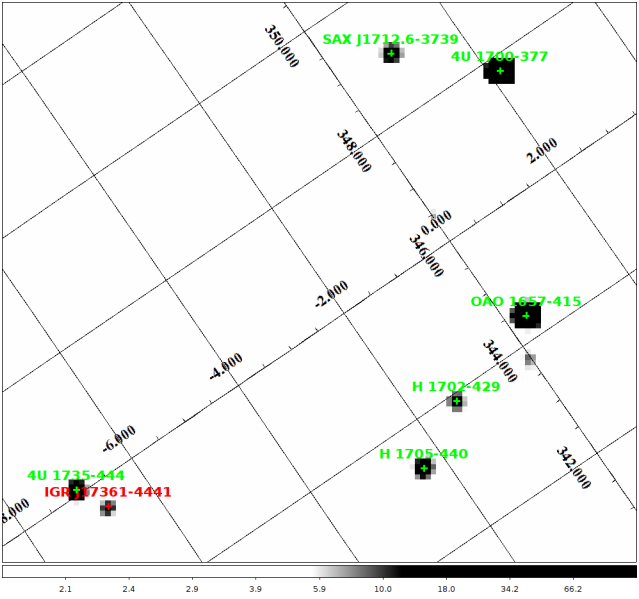}
  \caption{IBIS/ISGRI FOV around \igr\ (20-100~keV energy band). 
  We used data from rev. 1078 to 1083. The horizontal band on the bottom 
  of the image shows the colour-coded significances in the mosaic expressed 
  in units of $\sigma$.}   
  \label{fig:isgri}
\end{figure}

We analyzed all the available data for the IBIS/ISGRI \citep[20-150~keV,][]{lebrun03} 
and for the JEM-X telescope  \citep[5-20~keV,][]{lund03} 
onboard \inte\ that were performed in the 
direction of \igr\ from 2011 August 11 (rev. 1078)  
to 2011 August 29 (rev. 1083). 
Data analysis was carried out by using version 9.0 of the OSA 
software distributed by the ISDC \citep{courvoisier03}. 
In Fig.~\ref{fig:isgri}, we show the IBIS/ISGRI mosaicked significance map around 
\igr\ in the 20-100~keV energy band (total exposure time $\sim$500~ks).    
In this mosaic, \igr\ is detected at a significance level of 15~$\sigma$ 
at an average flux of 1.2$\times$10$^{-10}$~erg/cm$^{2}$/s. 
The corresponding IBIS/ISGRI spectrum is discussed later in Sect.~\ref{sec:swift}.  
From rev. 1078 to 1083, \igr\ was inside the field of view (FOV) of the X-ray monitor  
onboard \inte,\ JEM-X,  only for a total effective exposure time of 17~ks and was not detected. 
We derived a 3~$\sigma$ upper limit to the 3-35~keV X-ray flux from \igr\ of 3.8~mCrab\footnote{Calculated by using the 
calibration observation of the Crab in satellite rev. 967 (from 2010 September 13 to 15)} 
(corresponding to 1.1$\times$10$^{-10}$~erg/cm$^2$/s). 
This is compatible with the value expected based on the broad-band spectral properties of the 
source (see Sect.~\ref{sec:swift}).  

\subsection{ \rxte\ }
\label{sec:rxte}

\begin{figure}
\centering 
\includegraphics[scale=0.44]{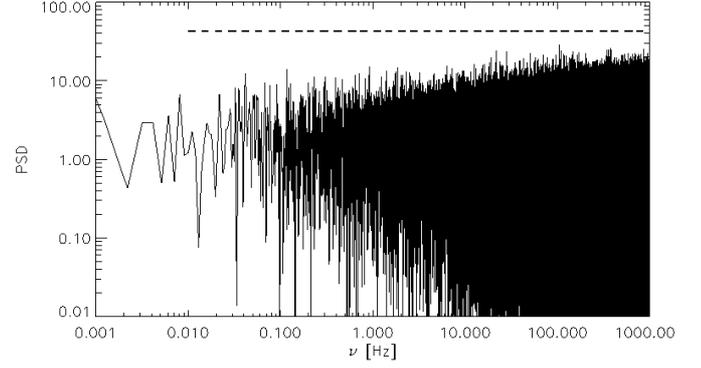}
\caption{Power spectral density obtained from the PCA events of PCU 1,2, and 4 
(we used a single time interval with bins of 0.49\,ms). The dashed line represents the 
$3\sigma$ limit for detection of coherent signals.}
\label{fig:psd} 
\end{figure}

A follow-up observation of \igr\ was performed with the PCA onboard \rxte\ 
\citep{rxte} on 2011 August 17 (total exposure time 1.6~ks). 
The pointing was carried out with an offset of $\sim$0.53 degrees from 
the nominal position of \igr\ to limit\footnote{By using archival PCA
observations, we estimated an average on-axis count-rate over three active PCUs of $\sim1400$\,cts/s for 4U\,1735$-444$.
Therefore, this source might contribute up to $\sim$25\% of the signal from the observation of \igr,\ 
which has a background-subtracted count-rate of $\sim10$\,cts/s.}
the contamination from the bright nearby low mass X-ray binary 4U\,1735$-444$. 
The good time intervals were based on a maximum offset from the source of 
0.55 degrees, a minimum elevation of ten degrees over the Earth limb, an electron ratio lower than 0.2, 
and at least ten minutes having passed since the SAA passage \citep[see also, e.g.][]{ferrigno11}. 
We analyzed the PCA event mode data in the GoodXenon configuration, which has a time resolution 
of $0.96\mu s$ and selected all events from the PCU 1, 2, and 4 (PCU0 suffered a breakdown 
during the observation). The event arrival times were all barycenter-corrected to the Solar 
System center-of-mass using the available tool {\sc faxbary} and the most-accurately determined 
source position in \chan\ data (see Sect.~\ref{sec:intro}).  
We constructed a power density spectrum after accumulating 
light curves in bins of 0.49\,ms, so that frequencies of up to $\sim$1\,kHz could be explored when searching for 
coherent signals (see Fig.~\ref{fig:psd}). To maximize the frequency resolution, we used a single 
interval to compute the fast Fourier transform (FFT). 
We did not find any significant peak at a $3\sigma$ c.l. \citep[we followed the approach described in][]{vaughan94}. 
The high background level of the PCA observation and the contamination from 4U\,1735$-444$  
did not permit an accurate estimation of the corresponding pulsed fraction upper limit. 

\subsection{ \swift\ }
\label{sec:swift}

\begin{figure}
\centering 
\includegraphics[scale=0.3,angle=-90]{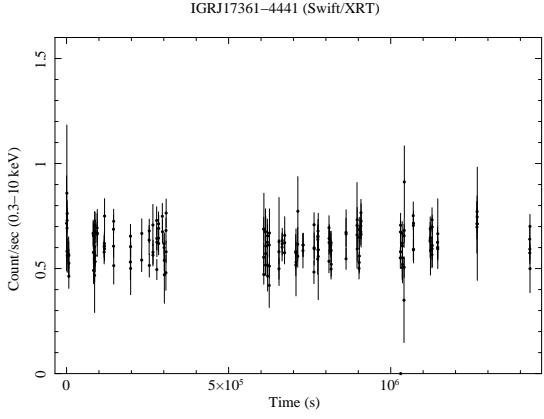}
\caption{Lightcurve of \igr\ as observed by \swift\,/XRT (0.3-10~keV). 
The bin time is 250~s and the start time is 2011 
August 16 at 17:01 (UTC).}
\label{fig:lcurve} 
\end{figure}

\begin{figure}
\centering 
\includegraphics[scale=0.30,angle=-90]{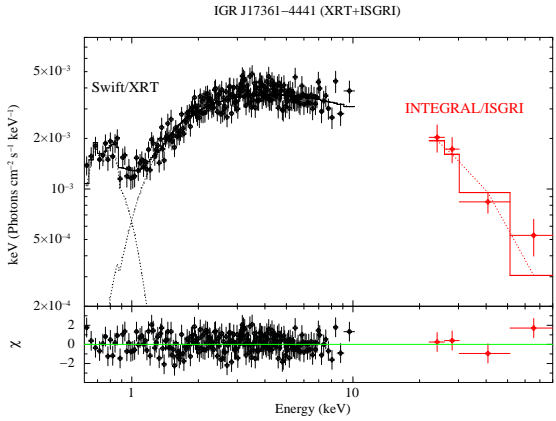}
\caption{Broad-band spectrum of \igr\ obtained by using data from 
\swift\,/XRT and \inte\,/ISGRI.  
The best-fit model (absorbed cut-off PL plus a BB component, 
see text for details) and the residuals from this fit are also shown.}    
\label{fig:spectrumtotal} 
\end{figure}

\swift\,/XRT monitored the outburst of \igr\ starting from 2011 August 16 for 
about 18~days (the last observation considered here was carried out on 2011 September 2 at 
06:10 UTC). We processed all the \swift\,/XRT data with 
standard procedures \citep{burrows05} and the latest calibration files available 
(caldb v. 20110725). All observations were performed in photon-counting mode 
(PC, time resolution 2.5~s). The data analysis technique that we used is the same as that 
described in \citet{bozzo09}.  
All the XRT spectra extracted from the available observations 
could be well accurately reproduced by using an absorbed power-law model with 
$N_{\rm H}$$\simeq$(0.2-0.5)$\times$10$^{22}$~cm$^{-2}$ 
and $\Gamma$$\sim$0.5-0.9. During this period, the source displayed a virtually constant 
X-ray flux of (4.5-4.8)$\times$10$^{-11}$~erg/cm$^2$/s (1-10 keV not corrected for absorption). 
The XRT lightcurve of the outburst is reported in Fig.~\ref{fig:lcurve}.  
By using the \swift\,/XRT on-line analysis tool\footnote{See http://www.swift.ac.uk/user\_objects/; see also 
\citet{evans09}.}  with all the XRT data available in the direction of NGC6388, we also determined the 
new XRT position of \igr\ to be  RA=17:36:17.27, Dec=-44:44:07.0, with an    
associated uncertainty of 1.9~arcsec (90\% c.l.; see Fig.~\ref{fig:positions}). 
This is consistent with the position reported previously by \citet{ferrigno11b} and the refined \chan\ position 
reported by \citet{pooley11}.    
We studied the broad-band spectral properties of the X-ray emission 
from \igr\ by combining XRT data carried out before 2011 August 29  
with those obtained during the same period with ISGRI (rev. 1078 to 1083). 
The joint XRT+ISGRI spectrum is shown in Fig.~\ref{fig:spectrumtotal}.  We fit the two spectra 
together by using an absorbed power-law (PL) model with a cut-off at high energy ({\sc cutoffpl} in {\sc Xspec}). 
A blackbody component (BB) was required to take into account the residuals at the lower energies ($\lesssim$2~keV). 
A normalization constant (fixed to 1 for the XRT spectrum) was included in the fit to account 
for the intercalibration between the different instruments and the variability (if any) of the source. 
From the fit ($\chi_{\rm red}^2$/d.o.f.=0.95/269), we estimated that 
$N_{\rm H}$=0.8$\pm$0.1~erg/cm$^2/s, \Gamma$=1.0$\pm$0.1, E$_{\rm cut}$=24$^{+20}_{-9}$~keV, 
$kT_{\rm BB}$=0.076$\pm$0.005~keV, and R$_{\rm BB}$=1.4$^{+0.8}_{-0.5}$$\times$10$^3$~d$_{10~kpc}$~km 
(where E$_{\rm cut}$ is the cut-off energy, $d_{10~kpc}$ is the source distance in units of 10~kpc, and all the 
uncertainties are given at the 90\% c.l.). 
The normalization constant used for ISGRI turned out to be C=1.2$_{-0.5}^{+0.8}$ and no significant changes 
in the spectral fit parameters were observed by fixing $C$=1. The source average X-ray flux was 
(4.6$^{+0.1}_{-0.5}$)$\times$10$^{-11}$~erg/cm$^2$/s in the 1-10~keV energy band and 
7.8$^{+0.8}_{-3.8}$$\times$10$^{-11}$~erg/cm$^2$/s in the 20-100~keV energy band. 
Compatible results were obtained by fitting the \rxte\,/PCA spectrum with the model described above. 
The 3-15 keV X-ray flux measured from the PCA was 6.7$^{+0.1}_{-3.4}$$\times$10$^{-11}$~erg/cm$^2$/s, i.e.  
slightly higher than that extrapolated from the XRT energy band. Part of this discrepancy might be due to  
the limited time range spanned by the PCA data with respect to XRT and ISGRI, and possible 
contamination of the nearby bright persistent low mass X-ray binary 
4U\,1735-444 (located $\sim$0.5~deg away from \igr,\ see also Sect.~\ref{sec:rxte}).  

\subsection{ {\sc ATCA} }
\label{sec:atca}

To hel us establish the nature of \igr,\ we also obtained on 2011 August 25   
a 8~h-long target of opportunity observation (ToO) with the Australia Telescope 
Compact Array \citep[ATCA;][]{wilson11} in configuration 6D (baseline is 6 km). 
The observation was conducted with a 2 GHz bandwidth centered around 5.5~GHz, 9~GHz, 
17~GHz, and~19 GHz. The total integration time was about 5.3~h for the two lower 
frequencies (observed simultaneously) and 2.7~h for the higher frequencies 
(observed simultaneously). 
Natural-weighted images were made from the ATCA data, with FOVs (FWHM) of 8.5, 5.2, 
2.8, and 2.5 arcmin, and synthesized beam sizes (FWHM) of 4.5$\times$1.8, 2.7$\times$1.1, 
1.7$\times$0.6, and 1.5$\times$0.5 arcsec at 5.5, 9, 17, and 19~GHz, respectively. 
No significant radio emission was found around the X-ray position of 
\igr.\ The rms noise levels of the images were 14.1~$\mu$Jy, 19.0~$\mu$Jy, 
20.0~$\mu$Jy, and 21.2~$\mu$Jy for the 5.5~GHz, 9~GHz, 17~GHz, and 19~GHz 
frequencies, respectively. The rms noise level corresponding to the image in the 
5.5~Ghz frequency is a factor of $\sim$2.5 lower than that reported previously by 
\citet[][see also Sect.~\ref{sec:discussion}]{cseh10}.

\section{Discussion}
\label{sec:discussion}

The X-ray monitoring campaign triggered after the discovery of \igr\ with \inte,\ has enabled us 
to significantly improve the accuracy of the source position, and confirm its 
location within the core of the GC NGC6388 \citep[the COG of the GC 
has coordinates RA=17:36:17.23, Dec=-44:44:07.1; the estimated uncertainty 
is 0.3~arcsec;][see also Sect.~\ref{sec:swift}]{lanzoni07}. 
\begin{figure}
\centering 
\includegraphics[width=8.0cm]{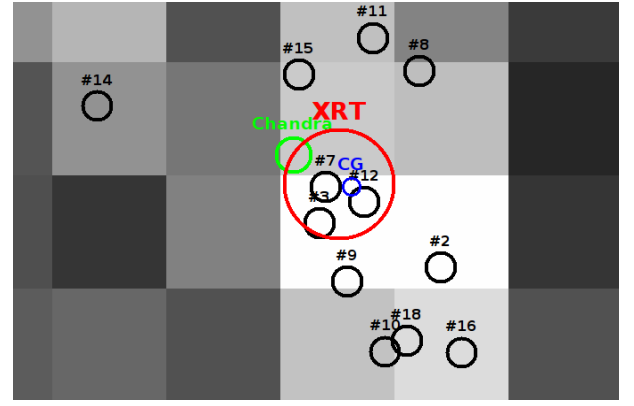}
\caption{Close view of the \chandra\ archival image (OBSID. 5055; 0.5-10 keV) around the core of NGC6388. 
In the image, black circles correspond to all the \chandra\ sources detected so 
far in quiescence by \citet[position uncertainty 0.5~arcsec;][]{cseh10}. The blue circle represents 
the COG of the GC (position uncertainty 0.3~arcsec). The improved XRT position determined 
in the present work is indicated with a red circle (see Sect.~\ref{sec:swift}), and the \chan\ position 
reported by \citet{pooley11} is represented with a green circle.} 
\label{fig:positions} 
\end{figure} 
The XRT data have helped us to demonstrate that the source remained at a virtually constant flux during the first 18~days 
after the onset of the outburst, and did not display any significant spectral variability. 
We have been able to fit the broad-band (0.5-100~keV) spectrum of the source reasonably well using a cut-off PL model 
($N_{\rm H}$$\simeq$0.8$\times$10$^{22}$~cm$^{-2}$, $\Gamma$$\simeq$0.7-1.0, and $E_{\rm cut}$$\simeq$25~keV) and 
displayed some hint of a  soft spectral component below $\sim$2~keV. 
The estimated flux in the 1-100~keV  energy band is 1.7$\times$10$^{-10}$~erg/cm$^2$/s, 
which corresponds to a luminosity of $L_{\rm X}$$\simeq$3.6$\times$10$^{36}$~erg/s 
at the distance of NGC6388 \citep[13.2~kpc;][]{dalessandro08}.  

The GC NGC6388 was previously observed with \swift,\ \xmm,\ and \chan,\ because it is 
believed to host an IMBH in its COG \citep[see also,][and references therein]{nucita08}. 
To date, the deepest X-ray observations of the GC were carried out 
with \chandra\ and reported by \citet{cseh10}. These authors found a number of sources located close 
to the core of the GC (sources \#3, \#7, and \#12), with spectral properties similar to those of 
the X-ray binaries\footnote{Source \#3 appeared to be rather soft and possibly compatible with being a 
cataclysmic variable. We do not discuss here this hypothesis for \igr\ as its X-ray luminosity seems 
to be far too high with respect to that typically observed from these objects \citep[see e.g.,][]{heinke08}.}. 
and averaged quiescent luminosities of $L_{\rm X}$$\simeq$10$^{32}$-10$^{33}$~erg/s. 
The relatively small uncertainty in the position of \igr,\ as determined thanks to the  
\chan\ observation of the source carried out by \citet{pooley11}, suggests that 
the newly discovered \inte\ transient is associated with neither the objects detected 
previously in quiescence by \citet{cseh10}, nor the COG of NGC6388 (see Fig.~\ref{fig:positions}). 
The quiescent luminosity of \igr\ can thus be estimated at $\sim$10$^{31}$~erg/s \citep{pooley11}, and  
is $\sim$10$^{5}$ times lower than it was at outburst. 
The radio observation discussed in Sect.~\ref{sec:atca}, showed that the large increase in the X-ray 
luminosity from \igr\ in outburst was not followed by any re-brightening in the radio domain. 
If we were to use our results obtained at 5.5~GHz and assume a distance to the source of 13.2~kpc, 
the rms level of 14.1~ $\mu$Jy derived in Sect.~\ref{sec:atca} 
from the ATCA data would translate into a 3$\sigma$ upper limit on the radio luminosity from \igr\ 
of $L_{\rm R}$$<$5$\times$10$^{28}$~erg/s \citep[we adopted the same calculation procedure as in][see their Sect.~4]{cseh10}.  
Assuming that the empirical relation between 
the mass of a BH and its radio and X-ray luminosity \citep[the so-called ``black-hole fundamental 
plane''][]{merloni03, kording06} applies also in the present case, 
the measured values of $L_{\rm X}$ and $L_{\rm R}$ do not seem to be compatible with 
\igr\ being a black-hole candidate (BHC) source, as the maximum allowed mass for this source 
would be far too low relative to expectations for these systems ($\gg$$M_{\sun}$). 

As the broad-band spectral properties of the source discussed in Sect.~\ref{sec:swift} are reminiscent 
of what is observed for some X-ray binaries hosting accreting neutron stars  
\citep[NSXRBs; see e.g.,][for a review]{psaltis04},  
we propose that an alternative possibility is that \igr\ belongs to this class of objects.  
Highly magnetized ($\sim$10$^{12}$~G) NSs in high mass X-ray binaries (HMXBs) 
usually display in outburst X-ray spectra comprising a relatively flat absorbed power-law ($\Gamma$$\simeq$0.5-1.0) 
component plus a cut-off at higher energies \citep[$\sim$10-50~keV, sometimes related 
to the strength of the NS magnetic field;][]{coburn02}. Soft excesses ($<$2~keV) are also observed from these 
sources in both quiescence and outburst, which are ascribed to either an accretion disk or a diffuse cloud 
around the compact object \citep[see e.g.,][]{hickox04}. Alternatively, a hot spot on the star can also give rise to a BB 
component in the spectrum with typical kT$\sim$0.3-1~keV. Given the results suggested by the XRT data analysis 
($R_{\rm BB}$$\simeq$1000~km), the first case would probably be the most likely one for \igr.\  
However, at odds with typical HMXBs in outburst, the new transient source did 
not display any pronounced variability in the X-ray emission on time-scales of neither a few hundreds of 
seconds (as in wind accreting systems) nor days (as in disk accreting systems). The relatively young age of typical 
HMXBs (few 10$^{6}$~yr) compared to that expected for the sources hosted in a globular cluster 
would also help us to exclude this hypothesis \citep[the age of NGC6388 has been 
estimated to be around 10$^{10}$~yr;][]{moretti09}. For this reason, \igr\ might be more easily 
associated with a low mass X-ray binary (ages $\gtrsim$10$^{8}$~yr), even though these objects usually 
have softer spectra in the X-ray domain ($\Gamma$$\sim$1.5-2). The lack of any pulsation in the \rxte\ data 
could also not be used to firmly establish the presence of a NS in \igr.\ 
As suggested by \citet{wijnands11}, a different possibility is that \igr\ is a very faint X-ray transient   
\citep[VFXT;][]{wijnands06,campana09}. These objects have similarly hard spectra 
($\Gamma$$\sim$0.6-1), and might emit a relatively stable X-ray flux in outburst even for a few years 
\citep[see, e.g., the case of XMMU\,J174716.1–281048;][]{delsanto07}. However, these sources typically have a lower 
X-ray luminosity (in the range 10$^{34}$-few$\times$10$^{35}$~erg/s) that that measured for \igr.\  
Further observations of this new \inte\ transient with sensitive X-ray telescopes in the next few months 
might help us to understand the real nature of \igr.\

Finally, we note that, as \igr\ seems to be a new X-ray source in NGC6388 and not associated with the IMBH 
possibly residing at its center, the ATCA upper limit derived in 
Sect.~\ref{sec:atca} gives a significantly tighter constraint of the allowed mass for this object with respect to   
those reported previously. Following the discussion in \citet{cseh10}, we assume that the quiescent X-ray luminosity of 
the IMBH is that determined by these authors with \chandra\ and consider the updated $L_{\rm R}$ provided by 
our ATCA ToO observation. Inserting these values into 
Eq.2 of \citet{cseh10} and taking into account the intrinsic scatter in the fundamental plane relation, the newly 
derived upper limit to the IMBH mass is $M$$<$600~$M_{\sun}$. This is a factor of 2.5 smaller than that reported by 
\citet{cseh10}.

\begin{acknowledgements} 
We thank the Swift, RXTE and ATCA teams for the rapid scheduling 
of the ToO observations of \igr,\ and an anonymous referee for useful comments. 
TB acknowledges financial support from ASI (INAF-ASI I/009/10/0). 
AP acknowledges the support of the operating program of Regione Sardegna
(ESF 2007-2013), L.R.7/2007, ``Promotion of scientific
research and technological innovation in Sardinia''. This research has made use of the IGR Sources page maintained 
by J. Rodriguez \& A. Bodaghee (http://irfu.cea.fr/Sap/IGR-Sources/).
\end{acknowledgements}

\bibliographystyle{aa}
\bibliography{ngc6388}

\end{document}